\documentstyle[aps,preprint,eqsecnum]{revtex}

\makeatletter
                \def\@preprint{}
                \def\preprint#1#2#3#4{%
                \ifpreprintsty
                \def\@preprint{
                \noindent \hbox{#1}\hfill\hbox{#2}\\
                \hbox{#3}\hfill\hbox{#4} \vskip 8pt}%
                \fi
                }
                \def\references{%
                \ifpreprintsty
                \vspace{2cm}
                \hbox to\hsize{\hss\large \refname\hss}%
                \else
                \vskip24pt
                \hrule width\hsize\relax
                \vskip 1.6cm
                \fi
                \list{\@biblabel{\arabic{enumiv}}}%
                {\labelwidth\WidestRefLabelThusFar  \labelsep4pt %
                \leftmargin\labelwidth %
                \advance\leftmargin\labelsep %
                \ifdim\baselinestretch pt>1 pt %
                \parsep  4pt\relax %
                \else %
                \parsep  0pt\relax %
                \fi
                \itemsep\parsep %
                \usecounter{enumiv}%
                \let\p@enumiv\@empty
                \def\theenumiv{\arabic{enumiv}}%
                }%
                \let\newblock\relax %
                \sloppy\clubpenalty4000\widowpenalty4000
                \sfcode`\.=1000\relax
                \ifpreprintsty\else\small\fi
                }

                \def\endreferences{%
                \def\@noitemerr
                {\@warning{Empty `thebibliography' environment}}%
                \endlist     \let\@SetMaxRefLabel\@gobble
                }
\makeatother

\begin{document}

\preprint{hep-ph/9511356}{PITHA 95/29}{November, 1995}{NTZ 28/95}

\title{
        Extraction of the pion distribution amplitude from \\
        polarized muon pair production
        }
\author{
        A. Brandenburg
        }
\address{
        Institut f\"ur Theoretische Physik, Physikzentrum,
        RWTH, 52056 Aachen, Germany
        }
\author{
        D. M\"uller
        }
\address{
        Institut f\"ur Theoretische Physik, Universit\"at Leipzig,
        04109 Leipzig, Germany
        }
\author{
        O. V. Teryaev
        }
        \address{
        Bogoliubov Laboratory of Theoretical Physics, \\
        Joint Institute for Nuclear Research, 141980 Dubna, Russia
        }

\maketitle{}

\begin{abstract}
We consider the production of muon pairs from the scattering
of pions on longitudinally polarized protons.
We calculate the cross section and the single spin asymmetry for this
process, taking into account pion bound state effects.
We work in the kinematic region where the photon has a large longitudinal
momentum fraction, which allows us to treat the bound state problem
perturbatively.
Our predictions are directly proportional to the pion distribution
amplitude.
A measurement of the polarized Drell-Yan cross section
thus allows the determination of the shape of the pion distribution
amplitude.
\end{abstract}


\narrowtext

\section{Introduction}

Spin effects are known
to provide sensitive tests for the underlying particle theories
in many cases.
One may recall the spin asymmetries in pion-nucleon scattering
which cannot be described in the simplest
pole approximation of the Regge theory and require to take into
account cuts.
In the case of QCD the so-called EMC Spin Crisis
was a signal of a breakdown of the naive parton model and
of the importance of the incorporation of
such a subtle field theory effect as the axial anomaly \cite{AEL}.

A less popular, but also extremely interesting example is the problem of
single transverse spin asymmetries in reactions
mediated by strong interactions.
As these effects require a mass parameter
and an imaginary part in the scattering amplitude,
they are proportional to $m_q \alpha_s$ in the
'naive' perturbative approach.
However, a careful application of QCD
factorization at twist-3 level removes both of these small parameters
\cite{ET85}. The collective gluon field of the hadron is shifting
the current quark mass $m_q$ to a mass parameter of the order
of the hadron mass and, simultaneously,
provides the imaginary part.
Technically, the latter appears when the pole of the quark (or gluon \cite
{StQiu}) propagator is integrated over the light-cone momentum fraction.

A similar imaginary part is present in the QCD higher twist contributions
to dilepton production in pion-nucleon scattering, due to the
integration over the light-cone momentum fraction of the quark in the pion.
This effect leads---in the case of unpolarized nucleon targets---
to a significant contribution to the dilepton angular distribution
\cite{BraBroKhoMue94}, which is
sensitive to different ans\"atze for the pion distribution function.

It is interesting to look for single spin asymmetries for
which the imaginary part does not constitute a (large) correction,
but the whole contribution.
The natural candidate is the spin asymmetry  in
the production of dileptons from the scattering of pions
on a {\it longitudinally polarized} nucleon. This spin asymmetry
has been thoroughly studied in perturbative QCD
\cite{PirRal83,CarWil92}.
The analogous mesonic processes have been discussed in \cite{OT95}.
In the present work we study this single spin asymmetry, taking into
account pion bound state effects.

\section{Calculation of the hadronic tensor}

We will consider the inclusive production of muon pairs
from the scattering of a pion on a polarized proton
(the Drell-Yan process),
\begin{eqnarray}
\pi^-(P_\pi)+p(P_p,s_\ell) \to \gamma^*(Q,\epsilon)+X
\to \mu^+(q_+)+\mu^-(q_-)+ X, \label{react}
\end{eqnarray}
where $s_{\ell}$ denotes the degree of longitudinal
polarization of the proton, and all momenta refer
to the overall c.m. system. It is convenient to
write the differential cross section for (\ref{react})
in terms of a hadronic and a leptonic tensor,
\begin{eqnarray}
\label{crosssectiondef}
d\sigma = {1\over 2s} \left({1\over Q^2}\right)^2
{d^3q_+ \over (2\pi)^3 2q_+^0 } {d^3q_- \over (2\pi)^3
2q^0_- } W^{\mu\nu}(s_{\ell}) L_{\mu\nu},
\end{eqnarray}
where $s$ is the hadronic c.m. energy. The leptonic
tensor has the well-known form
\begin{eqnarray}
\label{leptonictensor}
L^{\mu\nu}= 4 e^2  \left(q_+^{\mu} q_-^{\nu}
 + q_-^{\mu} q_+^{\nu} - g^{\mu\nu} Q^2/2 \right).
\end{eqnarray}
Note that $L^{\mu\nu}$ is symmetric, since we only consider
the exchange of a virtual photon (not a $Z$ boson) and since
we sum over the polarizations of $\mu^{\pm}$.
\par The hadronic tensor $W^{\mu\nu}$ may be written as
\begin{eqnarray}
\label{hadronictensor}
W^{\mu\nu}(s_\ell) &=&
\ \sum_X
(2\pi)^4 \delta^4(Q + P_X - P_\pi - P_p)
 \nonumber\\
& &\langle \gamma(Q,\epsilon^{\mu}),X(P_X)\vert {\cal T}\vert
\pi^-(P_\pi),p(P_p,s_{\ell})\rangle
\nonumber\\
& & \langle \gamma(Q,\epsilon^{\nu}),X(P_X)\vert {\cal T}\vert
\pi^-(P_\pi),p(P_p,s_{\ell})\rangle^*.
\end{eqnarray}
Only the symmetric part of $W^{\mu\nu}$ is relevant after
contraction with $L_{\mu\nu}$ of eq.\ (\ref{leptonictensor}).
Therefore $W^{\mu\nu}$ in the following is implicitly
assumed to be symmetrized.
\par The angular distribution of $\mu^+$ in (\ref{react})
may be parametrized as
\begin{eqnarray}
\label{angledist}
        {d\sigma\over dQ^2 dQ_T^2 dx_L d\cos\theta d\phi} &\propto&
        1+\lambda \cos^2\theta +\mu \sin 2\theta\cos\phi
        +{\nu \over 2}\sin^2\theta\cos 2\phi
\nonumber\\
        & &+\bar\mu \sin 2\theta \sin\phi + {\bar\nu \over 2}\sin^2\theta
        \sin 2\phi.
\end{eqnarray}
Here, $\theta$ and $\phi$ are angles defined in the muon
pair rest frame, and $\lambda,\ \mu,\ \nu,\ \bar\mu,\ \bar\nu
$ are angle independent coefficients. They depend on $Q^2$ (the
virtuality of the photon),
$Q_T^2$ (the squared transverse momentum of the photon
in the hadronic c.m.s.), and
on $x_L=2Q_L/\sqrt{s}$ (the longitudinal momentum fraction of
the photon). We take $x_L>0$ for a photon moving forward
with respect to the pion in the hadronic c.m. system.
The normalization of the cross section can be determined from
\begin{eqnarray}
\label{norm}
        {Q^2 d\sigma\over dQ^2 dQ_T^2 dx_L} = {e^2\over 96 (2\pi)^4}
        {(3+\lambda) N\over \sqrt{s} \sqrt{Q^2+Q_T^2+x_L^2 s/4} }
          .
\end{eqnarray}

The normalization $N$ and the angular coefficients are related
to the helicity amplitudes of the hadronic tensor:
\begin{eqnarray}
\label{angularcoef}
N &=&  W_T+W_L =
        \epsilon_{\mu}(+1) W^{\mu\nu}  \epsilon_{\nu}^\ast(+1)+
                \epsilon_{\mu}(0) W^{\mu\nu} \epsilon_{\nu}(0)
\nonumber\\
\lambda &=&  N^{-1}\left(W_T-W_L\right) = N^{-1}
         \large\{\epsilon_{\mu}(+1)  W^{\mu\nu} \epsilon_{\nu}^\ast(+1)-
                 \epsilon_{\mu}(0) W^{\mu\nu} \epsilon_{\nu}(0)\large\}
\nonumber\\
\mu &=&   N^{-1}W_{LT}= (\sqrt{2}N)^{-1}
         \large\{\epsilon_{\mu}(0) W^{\mu\nu} \epsilon_{\nu}(+1)+
\epsilon_{\mu}(0) W^{\mu\nu}\epsilon_{\nu}^\ast(+1)\large\}
\nonumber\\
\nu &=&   N^{-1}W_{TT}= N^{-1}
         \large\{\epsilon_{\mu}(+1) W^{\mu\nu} \epsilon_{\nu}^\ast(-1)+
        \epsilon_{\mu}^\ast(+1) W^{\mu\nu}\epsilon_{\nu}(-1)\large\}
\nonumber\\
\bar{\mu} &=&  N^{-1}\overline{W}_{LT}=i(\sqrt{2}N)^{-1}
        \large\{\epsilon_{\mu}(0) W^{\mu\nu} \epsilon_{\nu}(+1)-
        \epsilon_{\mu}(0) W^{\mu\nu}\epsilon_{\nu}^\ast(+1)\large\}
\nonumber\\
\bar{\nu} &=&  N^{-1} \overline{W}_{TT} = i N^{-1}
         \large\{\epsilon_{\mu}(+1) W^{\mu\nu}\epsilon_{\nu}^\ast(-1)-
        \epsilon_{\mu}^\ast(+1) W^{\mu\nu}\epsilon_{\nu}(-1)\large\},
\end{eqnarray}
where the polarization vectors
$\epsilon^\mu(\pm 1)=(0,\mp{\bf e}_x+i{\bf e}_y)$ and
$\epsilon^\mu(0) = (0,{\bf e}_z) $ are determined by specifying
the coordinate axes ${\bf e}_i\ (i=x,y,z)$ in the muon rest frame.

The coefficients $\bar{\mu}$ and $\bar{\nu}$ are nonzero only
in the polarized Drell-Yan process and are
induced  by absorptive parts in the scattering amplitude ${\cal T}$.
In the parton model, the leading contributions to $\bar{\mu}$ and $\bar{\nu}$
come from the interference of Born diagrams with the absorptive parts of
one-loop diagrams \cite{CarWil92}. The relation of the
single spin asymmetry
\begin{eqnarray}\label{asymm}
{\cal A}\equiv {d\sigma(s_{\ell}=+1)-d\sigma(s_{\ell}=-1)\over
d\sigma(s_{\ell}=+1)+d\sigma(s_{\ell}=-1)}
\end{eqnarray}
 to the angular coefficients is simply given by
\begin{eqnarray} \label{connection}
{\cal A}=
        {\bar{\mu}(s_{\ell}=+1)\sin 2\theta\sin\phi+
        {1\over 2}\bar{\nu}(s_{\ell}=+1)\sin^2\theta\sin 2\phi
\over
        1+\lambda \cos^2\theta +\mu \sin 2\theta\cos\phi+
        {1\over 2}\nu\sin^2\theta\cos 2\phi}.
\end{eqnarray}

For the unpolarized Drell-Yan process the parton model predictions
\cite{ChiBel86} fail to
describe the data on $\lambda,\ \mu$ and $\nu$ \cite{expunpol}.
In contrast, taking into account pion bound state effects yields reasonable
fits to the data \cite{BraBroKhoMue94}. An interesting feature of this
bound state model is a nonvanishing absorptive part in {\it leading} order
which is
proportional to the pion distribution amplitude $\varphi$.
This property directly
leads to nonvanishing  $\bar{\mu}$ and $\bar{\nu}$
which are strongly sensitive
to $\varphi$. In the following we will shortly describe
the salient features of the bound state model for the pion.\
Further details may be found, e.g., in
\cite{BroBerLep}, \cite{BraBroKhoMue94}.

The bound state problem may be treated perturbatively if
the momentum fraction of the quark from the pion is large,
which is the case for large $x_L$.
The dominant contribution to the Drell-Yan process comes from the
annihilation of a quark with an antiquark \cite{DreYan70}.
In the region of large $x_L$, the
diagrams of fig. 1a,b. then give the leading contributions.
We will explain the formalism for the annihilation of a $u$ quark
from the proton with the $\bar{u}$ quark from the pion; the other
possible annihilation process of a $\bar{d}$ quark from the proton
with the $d$ quark from the pion is treated in complete analogy
and its contribution will later be added incoherently.

In diagram 1a we see that the
 $\bar{u}$ quark propagator is far off-shell,
$p_{\bar{u}}^2=-Q_T^2/(1-x_{\bar{u}})$, if
the light cone momentum fraction of the quark
from the pion $x_{\bar{u}}\approx x_L$
is close to 1 \cite{BroBerLep}.
Thus, the pion can be resolved by a single hard gluon exchange \cite{BroLep80}.
The second diagram, fig. 1b, is required by gauge invariance.

The hadronic tensor $W^{\mu\nu}$ is obtained by a convolution
of the partonic tensor $w^{\mu\nu}$ with the corresponding
parton distribution functions for the polarized proton\footnote
{Here we take into account the correct partonic flux factor
which amounts to multiplying with $s/\hat{s}$, $\hat{s}=
(P_\pi + p_u)^2$.}.
The partonic tensor $w^{\mu\nu}$ is computed from the product
$M^{\mu}M^{*\nu}$. $M^{\mu}$ is the amplitude for
the reaction
\begin{eqnarray}
\label{partreact}
u^{\uparrow}+\pi^- \to \gamma^*(Q,\epsilon^\mu)+X,
\end{eqnarray}
where the arrow indicates the polarization of the parton.
This amplitude $M^{\mu}$ is obtained by convoluting the partonic amplitude
 $T^{\mu}(u^{\uparrow}+\bar{u}d \to \gamma^*+d)$ with the pion
distribution amplitude $\varphi(z,\tilde{Q}^2)$ \cite{BroLep80},
\begin{eqnarray}
\label{convdist}
M^{\mu}= {f_\pi \over 4N_c} \int_0^1 dz \ \varphi(z,\tilde{Q}^2) \ T^{\mu},
\qquad \int_0^1 dz \ \varphi(z,\tilde{Q}^2) = 1,
\end{eqnarray}
where $\tilde{Q}^2\sim Q_T^2/(1-x_{\bar{u}})$ is the cutoff for the
integration over soft momenta in the definition of $\varphi$,
$f_\pi \approx 133 {\rm \ MeV}$ is the pion decay constant, and $N_c=3$ is the
number of colors.
A detailed discussion of the scale setting problem for
exclusive processes is given in \cite{BroHun94}. Using the commensurate
scale relations in the calculation of (\ref{convdist}) would allow
to eliminate ambiguities caused by the factorization
and renormalization. However, this does not change the qualitative features
of our predictions and thus it will not be discussed in this paper.

The imaginary part in this model arises because the internal quark line of
fig. 1b can go on-shell. After integrating the hard scattering amplitude
$T^{\mu}$ according to equation (\ref{convdist}), we are left with a regular
amplitude $M^{\mu}$ which has an imaginary part proportional to the pion
distribution amplitude at the point where the quark propagator gets singular.

We are now ready to present our  analytic results.
We relegate the full result for the hadronic tensor to the appendix, and
confine ourselves here to a discussion of the
angular coefficients $\bar{\mu}$ and $\bar{\nu}$.
To present our results for the angular distribution
we choose the
Gottfried-Jackson frame where the ${\bf e}_z$ axis is taken to be the pion
direction in the muon rest frame and the ${\bf e}_x$ axis lies in
the $\pi^- P$
plane such that the proton momentum has a negative $x$ component. From
eqs.\ (\ref{angularcoef}),\ (\ref{hadronictensordecomposition}), and
(\ref{structurfunction}) we get the angular coefficients
\begin{eqnarray}
\label{mubarnubar}
\bar{\mu} &=&
        {-2  s_\ell \rho \tilde{x} F\pi\varphi(\tilde{x}) \over
        (1-\tilde{x})^2 \left[(F+{\rm Re\ } I(\tilde{x}))^2
        + \pi^2\varphi(\tilde{x})^2\right] + \rho^2\tilde{x}^2 F^2(4+\rho^2)}
\nonumber\\ & &\qquad \times \
        {{4\over 9} \Delta q_u^v(x_p) + {4\over 9} \Delta q_u^s(x_p) +
        {1\over 9} \Delta q_d^s(x_p)\over
        {4\over 9} q_u^v(x_p) + {4\over 9} q_u^s(x_p) + {1\over 9}
        q_d^s(x_p)},
\nonumber\\
\bar{\nu} &=& 2 \rho \bar{\mu}.
\end{eqnarray}
Here, $\rho=Q_T/Q$ and $\tilde{x}$ is a function of $x_L, Q^2/s$\ and $\rho$
\begin{eqnarray}
\label{deftildex}
\tilde{x} \equiv {x_{\bar{u}}\over 1+\rho^2}={1\over 2}
        {x_L +
\sqrt{x_L^2 +4Q^2 s^{-1}(1+\rho^2)}\over 1+\rho^2}.
\end{eqnarray}
Furthermore,
\begin{eqnarray}
        F&=&\int_0^1 dz{\varphi(z,\tilde{Q}^2)\over z},\nonumber \\
        I(\tilde{x})&=&\int_0^1 dz
        {\varphi(z,\tilde{Q}^2)\over z(z+\tilde{x}-1+i\epsilon)}
\end{eqnarray}
denote integrals over the pion distribution amplitude.
Finally, $q_i^j(x_p)$ and
$\Delta q_i^j(x_p)$
are the unpolarized and polarized valence and sea quark distribution
functions evaluated at the point
$x_p\approx Q^2 /(s \tilde{x})$.
The results  for $\lambda,\ \mu,\ \mbox{and}\ \nu$ within the
 pion bound state model
are given elsewhere
\cite{BraBroKhoMue94}.

\section{Numerical results}
The numerical results for the angular coefficient functions $\bar{\mu}$ and
$\bar{\nu}$, eq. (\ref{mubarnubar}), depend on $\tilde{x}$, $\rho$, the
pion distribution amplitude and on the ratio of the polarized and
unpolarized quark distribution functions of the proton. Whereas the
unpolarized quark distributions are known quite well, the polarized
ones still contain large uncertainties. For our results we choose the
parametrizations given in \cite{GehSti95} at $Q^2=4 {\rm \ GeV}^2$.

In fig.\ 2 we present our results for several  distribution amplitudes
$\varphi$
that are quite different in shape. Fig.\ 2a shows our
choices for $\varphi$: The solid line
represents the two-humped function \cite{CzeZhi80} which gave a good fit
to the data on $\lambda,\ \mu$ and $\nu$  \cite{BraBroKhoMue94}.
The effective evolution parameter is set to
$\tilde{Q}^2 \sim 4 {\rm \ GeV}^2$.
Since the predictions are also very sensitive to the behavior of the
distribution amplitude in the endpoint region, we
choose two different parametrizations \cite{MikhRad}
for the convex distribution amplitude
$\varphi(z) = z^a(1-z)^a/{\rm B}(a+1,a+1)$.
The dashed line in fig. 2a represents the asymptotical amplitude, i.e., $a=1$,
and the dotted line shows a narrow distribution amplitude with $a=10$.

In fig.\ 2b  the moment
$\int \sin 2\theta \sin\phi d\sigma(s_{\ell}=+1) \propto N \bar{\mu}$ is
plotted in arbitrary units versus $x_L$ for the values
$s=20^2 {\rm \ GeV}^2$, $Q=3$ GeV and $Q_T=0.9$ GeV. It can be seen that
different shapes of the distribution
amplitude could be distinguished by a measurement of this quantity,
which is proportional to the pion distribution amplitude evaluated
at $\tilde{x}\approx x_L$ (cf. the numerator of (\ref{mubarnubar})).
It is also demonstrated that narrow
distribution amplitudes give a moment which vanishes in the large $x_L$
region.

In  figs.\ 2c and 2d the angular coefficient
$\bar{\mu}(s_{\ell}=+1)$ is plotted versus $x_L$ for two different choices
of $Q_T$ and the same values for $s$ and $Q$ as in fig. 2b. In fig. 2c
$Q_T$ was set again to $0.9$ GeV, which gives the moderate value $\rho=0.3$.
It is demonstrated that
the two-humped form for $\varphi$ induces a minimum at
$x_L \sim 0.6$ which, however, vanishes for the smaller value
$\rho=0.06$ used in fig. 2d. For very narrow
distribution amplitudes $\bar{\mu}$ is strongly suppressed. Since
$\bar{\nu}=2\rho\bar{\mu}$ (cf. eq. (\ref{mubarnubar})) in the bound state
model, we do not show separate plots for this quantity.

Note that the effects of a variation of the pion distribution amplitude
in the spin-dependent and spin-averaged cross sections --
i.e. in the numerator and denominator
of (\ref{mubarnubar}) -- partially compensate each other. As a consequence,
the result for the two-humped distribution
amplitude appears in between
the results for the two convex distributions. The spin-dependent part
of the cross section is a more sensitive ``partonometer'', than
the dimensionless coefficients $\bar \mu, \bar \nu$. Although the former
quantity suffers from larger uncertainties due to higher order corrections,
, one may expect only a small
uncertainty for the determination of the shape of pion distribution
amplitude, if the $x_L$ dependence of the corrections is
weak.

Although the naive parton model is
at variance with the unpolarized angular distribution
\cite{expunpol}, we would like to present for completeness and comparison also
the predictions of this model for $\bar{\mu}$
and $\bar{\nu}$. We took the analytic results from \cite{CarWil92}
and again used the parton distribution functions of \cite{GehSti95}.
In the range $0.5 < x_L <1$ the parton model yields coefficients
$\bar{\mu}(s_{\ell}=+1)$ and $\bar{\nu}(s_{\ell}=+1)$ which are positive
and to a very good approximation independent of $x_L$.
For the same values used in fig2b and 2c, i.e. $s=20^2 {\rm \ GeV}^2,
Q=3 {\rm \ GeV}$ and $\rho = 0.3$  we get a value of
$\bar{\mu}(s_{\ell}=+1)\sim 0.12 \alpha_s \sim 0.036$.
This means that bound state effects
are roughly of the same magnitude but opposite in sign
compared to the parton model at
moderate values of $\rho$. For the smaller value $\rho = 0.06$
we find $\bar{\mu}(s_{\ell}=+1)\sim 0.034 \alpha_s \sim 0.01$, again
independent of $x_L$ and positive.
At such a small value of $\rho$ the bound state model predicts
 that $\bar{\mu}$ and $\bar{\nu}$ are
concentrated at large $x_L$ for not too narrow distribution amplitudes.
In the kinematical range considered, the relation between $\bar{\nu}$ and
$\bar{\mu}$ calculated within the parton model
takes a similar form as in the bound state model, namely,
$\bar{\nu} \sim 3 \rho \bar{\mu}$.

In conclusion we have shown that experiments on dimuon production from the
scattering of pions on polarized targets will give detailed information both
on the shape and on the endpoint behavior
of the pion distribution amplitude.

\acknowledgments
We would like to thank S. J. Brodsky, S. V. Mikhailov, J. Soffer and
A. V. Radyushkin for valuable discussions. A. B. and O. V. T. would like to
thank also the Naturwissenschaftlich-Theoretisches Zentrum and
the Graduiertenkolleg "Quantenfeldtheorie" at Leipzig University for
the kind hospitality during their stay, where substantial parts of this work
were done. D.M. was financially supported by the Deutsche
Forschungsgemeinschaft. O.V.T. was supported in part by Russian Foundation
for Fundamental Researches, International Science Foundation and INTAS.

\appendix
\section*{Result for the hadronic tensor}

Using the kinematic decomposition of the
(symmetric part of the) hadronic tensor
\begin{eqnarray}
\label{hadronictensordecomposition}
W^{\mu\nu} &=&
         -\tilde{g}^{\mu\nu} W_1
+ {\tilde{P}^\mu_\pi \tilde{P}^\nu_\pi\over Q^2} W_2
         + {\tilde{P}^\mu_p \tilde{P}^\nu_p \over Q^2} W_3
+ {(\tilde{P}^\mu_\pi \tilde{P}^\nu_p
+ \tilde{P}^\mu_p \tilde{P}^\nu_\pi)\over  Q^2} W_4
\nonumber\\
& & +{ \left(\epsilon^\mu_{\ \alpha\beta\gamma} \tilde{P}^\nu_\pi +
\epsilon^\nu_{\ \alpha\beta\gamma} \tilde{P}^\mu_\pi\right)
P^\alpha_\pi P^\beta_p Q^\gamma \over  Q^4} s_\ell W_5
\nonumber\\
& &+{ \left(\epsilon^\mu_{\ \alpha\beta\gamma} \tilde{P}^\nu_p +
\epsilon^\nu_{\ \alpha\beta\gamma} \tilde{P}^\mu_p\right)
P^\alpha_\pi P^\beta_p Q^\gamma\over  Q^4} s_\ell W_6,
\end{eqnarray}
where $\tilde{g}^{\mu\nu} = g^{\mu\nu} - Q^{\mu} Q^{\nu}/Q^2$ and
$\tilde{P}^{\mu}_i = \tilde{g}^{\mu}_{\ \nu} P_i^\nu$,
we obtain  from the bound state model
the following predictions for the structure
functions:
\begin{eqnarray}
\label{structurfunction}
W_1 &=& {n\over 4}
        \left\{\left(\left[1-\tilde{x}(1+\rho^2) \right]F +
        (1-\tilde{x}){\rm Re\ } I(\tilde{x})\right)^2+
        \pi^2 \varphi(\tilde{x})^2\right\} f\left(x_p\right)
\nonumber\\
W_2 &=& -n \tilde{x}^2 f\left(x_p\right)         \nonumber \\ \ & &
        \left\{(F+(1-\tilde{x}) {\rm Re\ } I(\tilde{x}))
        \left[(1-\tilde{x}(1+\rho^2)^2) F
         +(1-\tilde{x}) {\rm Re\ } I(\tilde{x}))\right]+
         \pi^2 \varphi(\tilde{x})^2\right\}
\nonumber\\
W_3 &=& {n x_p^2 \tilde{x}\over 1-\tilde{x} }
        \left[1-\tilde{x}(1+\rho^2)\right]^2 F
        (F+{\rm Re\ } I(\tilde{x})) f\left(x_p\right)
\nonumber\\
W_4 &=& -  {n x_p \tilde{x}^2 \over 1-\tilde{x} } \rho^2
        \left[1-\tilde{x}(1+\rho^2)\right]
        F(F+(1-\tilde{x}) {\rm Re\ } I(\tilde{x})) f\left(x_p\right)
\nonumber\\
W_5 &=&
        -{n x_p \tilde{x}^3 \over 1-\tilde{x} } (1+\rho^2)
        \left[1-\tilde{x}(1+\rho^2)\right]
        \pi\varphi(\tilde{x}) F \Delta f\left(x_p\right)
\nonumber\\
W_6 &=&
        {n x_p^2 \tilde{x}^2 \over (1-\tilde{x})^2 }
        \left[1-\tilde{x}(1+\rho^2)\right]^2
        \pi\varphi(\tilde{x}) F  \Delta f\left(x_p\right).
\end{eqnarray}
Here,
\begin{eqnarray}
\label{normalization}
n={32 \pi^3 e^2 \alpha_s^2 C_F^2 f_\pi^2 \over N_c^2 Q_T^4}\quad
\mbox{with\ } C_F={N_c^2-1\over 2N_c}
\end{eqnarray}
is a normalization factor, and
\begin{eqnarray}
\label{deffanddeltaf}
f(x_p)&=& {4\over 9} q_u^v(x_p) + {4 \over 9} q_u^s(x_p)
 + {1\over 9} q_d^s(x_p)
\nonumber\\
\Delta f(x_p)&=&{4 \over 9} \Delta q_u^v(x_p) + {4\over 9} \Delta
q_u^s(x_p) + {1\over 9} \Delta q_d^s(x_p),
\end{eqnarray}
are given by the unpolarized and polarized valence and sea quark distribution
functions.
The variable $\tilde{x}$ and $x_p$ as well as the integrals
$F$ and ${\rm Re}\ I$
containing the pion distribution amplitude are defined below
eq.\ (\ref{mubarnubar}).

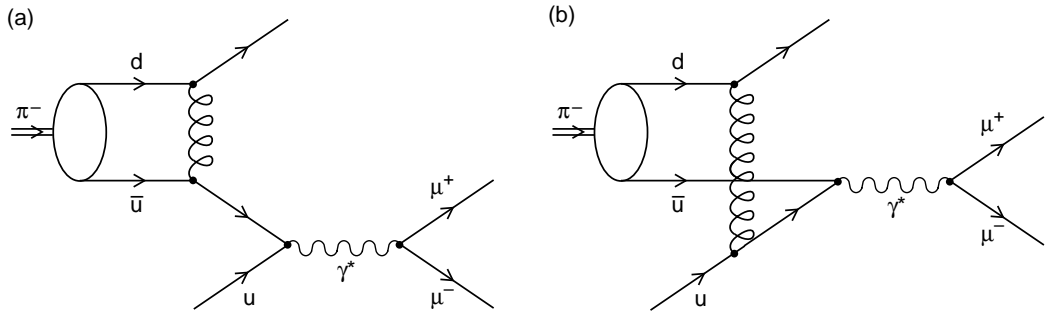
\begin{figure}
\caption{Leading contributions  to the amplitude for the Drell-Yan process at
large $x_L$.}
\label{fig1}
\end{figure}

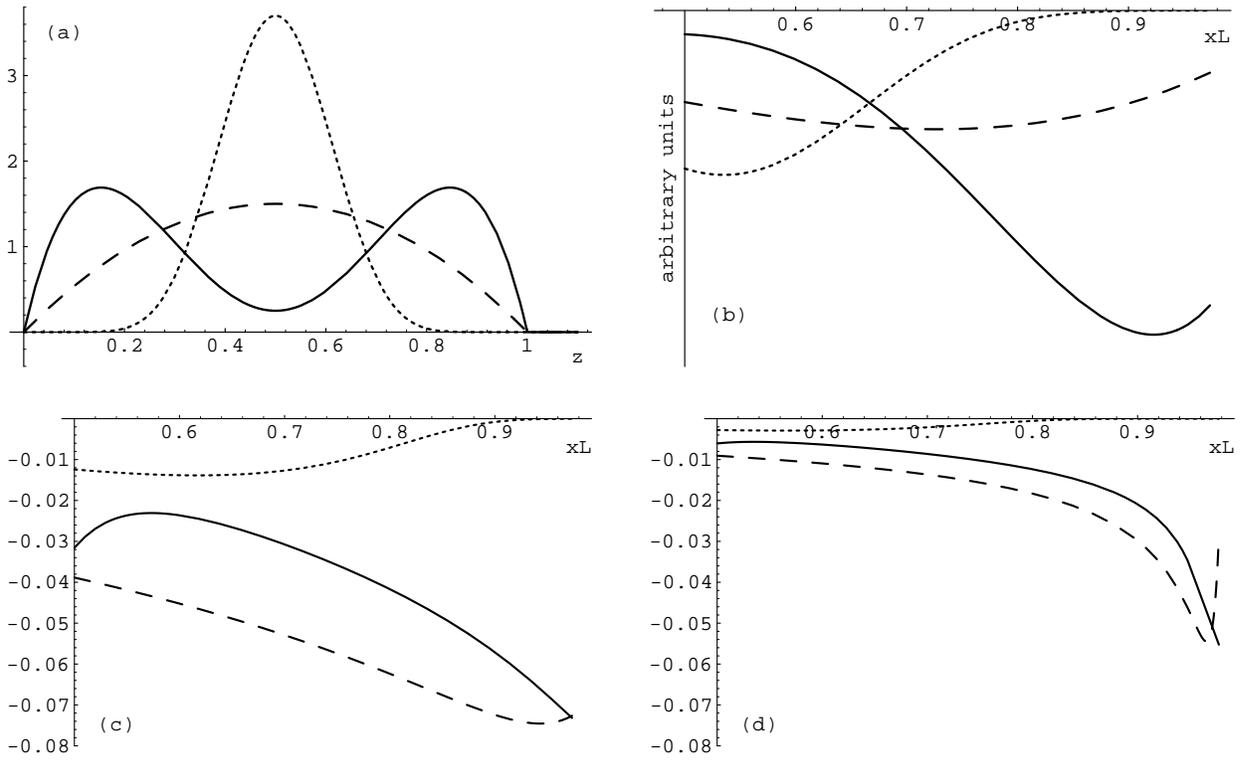
\begin{figure}
\caption{In (a) we show different pion distribution
amplitudes $\varphi(z)$. Solid line:
two-humped function with an
effective evolution parameter of $\tilde{Q}^2 \sim 4 {\rm \ GeV}^2$;
dashed line: $\varphi(z) = z^a(1-z)^a/{\rm B}(a+1,a+1)$ with $a=1$;
dotted line: same with $a=10$.
In (b) the moment $\int \sin 2\theta \sin\phi d\sigma(s_{\ell}=1)$ is
plotted versus $x_L$ for the values $s=20^2 {\rm \ GeV}^2$, $Q=3$ GeV and
$Q_T=0.9$ GeV.
In  (c) and (d) the angular coefficient $\bar{\mu}(s_{\ell}=1)$ is plotted
versus $x_L$ for the two different values $Q_T/Q=0.3$ and $Q_T/Q=0.06$,
respectively, and for the same values for $s$ and $Q$ as in (b). In (b-d)
we use the same choices for $\varphi(z)$ as in (a).}
\label{fig2}
\end{figure}

\end{document}